# Graphene-Enabled Terahertz Metamaterial Absorber for Ultra-Wide Refractive Index Sensing

Osama Haramine Sinan[a*], Roktim Barua[a], Nipa Dhar[a], Muhammad Asad Rahman[a]

[a]Department of Electrical and Electronic Engineering,
Chittagong University of Engineering and Technology (CUET),
Chattogram-4349, Bangladesh

*Corresponding author

**Abstract:**

This study introduces a graphene-assisted terahertz metamaterial absorber operating at a resonance frequency of 8.436 THz. The unit cell incorporates a graphene-based split-ring resonator (SRR) configuration with additional auxiliary splits on the top layer. The structure achieves near-perfect absorption of 99.99% under TM polarization, maintaining absorption above 90% for incident angles up to 75°. In addition, the proposed design remains polarization-insensitive up to 50°. The sensor demonstrates a figure of merit (FOM) of 4.22 $RIU^{-1}$ and achieves a maximum Q-factor of 20.98, corresponding to a full width at half maximum (FWHM) of 402 GHz. An equivalent circuit model is also developed to verify the resonance behavior. Due to the strong dependence of the resonant frequency on the surrounding refractive index (RI), the absorber is applied as an RI-based sensor. It exhibits a high sensitivity of 1698 GHz/RIU across a wide refractive index range from 1.0 to 2.0. The proposed design holds promise for sensing applications involving gases, oils, solvents, polymers, and high-index dielectric materials, including the biomedical refractive index window (1.30–1.39) for label-free biosensing.

## 1. Introduction

Metamaterials (MMs) are artificially engineered media composed of subwavelength unit cells that enable effective electromagnetic properties (ε, μ) not found in natural materials, including, in some cases, a negative refractive index (RI) as predicted by Veselago and experimentally realized in structures exhibiting simultaneously negative permittivity and permeability [1–4]. Building on these foundational concepts, optical and terahertz (THz) metamaterials have demonstrated negative refraction, strong near-field confinement, spectral selectivity, and tailored absorption/emission characteristics, supporting applications in imaging, communications, and sensing [5,6]. In the THz range (0.1–10 THz), where many molecular vibrational and phonon resonances lie, MMs offer compact, highly resonant components capable of strong interaction with very small analyte volumes while maintaining non-ionizing characteristics, making them well suited for chemical, industrial, environmental, and biomedical diagnostics.

Within this field, perfect metamaterial absorbers (PMAs) are especially attractive for THz sensing. Their metal–dielectric–metal (or metasurface–spacer–ground) configuration can be engineered to achieve free-space impedance matching, enabling near-complete absorption and strong field confinement within a subwavelength cavity [7,8]. Changes in the surrounding refractive index (RI) then lead to measurable spectral shifts, enabling label-free detection.

Nevertheless, most traditional metal-based PMAs lack post-fabrication tunability, limiting their adaptability to diverse analytes without a geometric redesign. While phase-change media (e.g., $VO_2$), doped semiconductors, and transparent conductive oxides enable limited reconfiguration, they often require thicker films or provide modest tunability within the THz spectrum [9]. Graphene, by contrast, is an atomically thin, electrically tunable plasmonic conductor whose surface conductivity—described by the Kubo formalism—can be dynamically adjusted via chemical potential (electrostatic gating, doping, or optical excitation) [10–12]. Patterned graphene supports ultra-confined THz plasmons, enhancing field overlap with analytes and improving RI sensitivity, with studies demonstrating competitive or superior performance compared to gold-based resonators [13–15]. In addition, π–π interactions and charge-transfer effects at the graphene–analyte interface can further alter graphene conductivity, enabling enhanced detection at very low surface concentrations [16–18].

Motivated by these characteristics, a graphene-based THz PMA is developed as an RI sensor spanning n = 1.0–2.0. The design leverages impedance-matched perfect absorption and strongly confined THz modes to channel incident energy into the resonant cavity, maximizing absorptance when the effective impedance approximates free space [19]. Electrical tunability via graphene's chemical potential allows resonance positioning, fine control, and stabilization as the surrounding RI changes, while graphene's high kinetic inductance supports deep subwavelength confinement and improved light–matter interaction [20,21]. These principles align with experimental demonstrations showing tunable perfect absorption and analyte-induced spectral modulation in graphene metasurfaces, confirming their suitability for high-FOM RI sensing [16,21]. The wide RI operating range covers gases, aqueous and organic solutions, polymers, oils, and high-index dielectrics, while naturally including biomedical targets within the n = 1.30–1.39 window [19,22,23].

## 2. Structure and Design:

A schematic of the THz metamaterial absorber designed for refractive index sensing is shown in Fig. 1. The unit cell is square with dimensions a = 10 μm, and the patterned resonator occupies b = 7 μm. It features a primary split-gap of g = 0.5 μm, an auxiliary slit of p = 0.1 μm for added capacitive loading, and a trace width of m = 0.5 μm. The device follows a standard absorber configuration: a patterned graphene layer on the sensing side with thickness t_gp = 0.08 μm, an FR-4 dielectric spacer of thickness t_s = 1.25 μm, and a continuous gold ground plane of thickness t_Au = 0.09 μm, which eliminates transmission.

The S-shaped graphene resonator behaves as an LC circuit, where inductance is provided by the current flowing along its arms, and capacitance is concentrated across the engineered gaps, producing localized electric-field hot spots critical for sensing. The central gap g forms the primary capacitive region, while the secondary slit p modifies current flow to introduce additional capacitance, optimize bandwidth, enhance polarization and angular robustness, and generate an extra hot spot.

Full-wave 3D electromagnetic simulations are conducted using periodic (Floquet) boundary conditions and normal plane-wave excitation. Material parameters for the graphene layer are defined using macros with T = 300 K, Fermi level E_F = 0.3 eV, relaxation time = 0.2 ps, and thickness = 0.1 nm. The overall design target is to realize a perfect-absorption resonance suitable for RI sensing. A split-gap configuration is chosen for its well-defined LC characteristics and strong field localization, while the auxiliary slit is introduced to further enhance capacitive behavior and enable multi-parameter tuning, with structural near-symmetry retained to achieve polarization-insensitive performance.

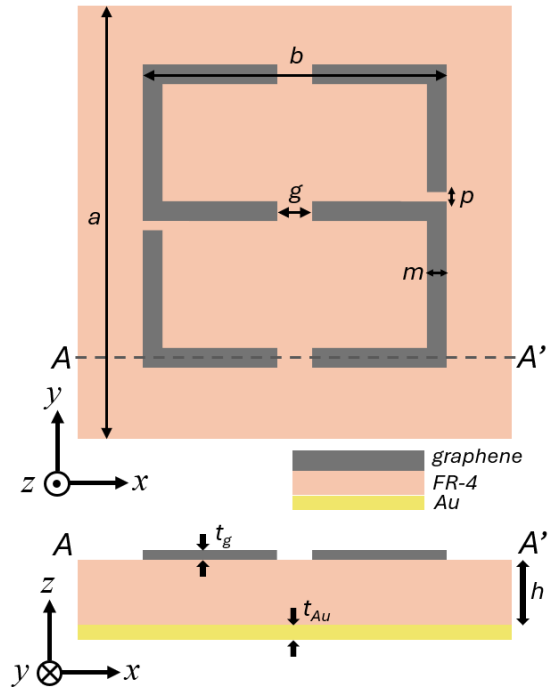

Fig 1: Structure of the proposed metamaterial absorber

### 3. Fabrication Process Overview:

The proposed THz metamaterial absorber is composed of a patterned graphene metasurface, an FR-4 dielectric layer, and a continuous gold (Au) ground plane. Its fabrication can be carried out using standard thin-film and PCB-compatible microfabrication processes, including vapor deposition, graphene transfer, and lithographic patterning. Additionally, cost-effective methods such as direct-write or printing-based techniques may be utilized. All fabrication steps are designed to comply with FR-4 thermal limits and follow typical industrial processing standards [24].

#### 3.1. Standard Thin-Film Fabrication:

In this approach, the FR-4 substrate functions as both the dielectric spacer and the mechanical support layer. A thin metal adhesion layer (Ti/Cr, 5–10 nm) is first deposited, followed by a gold coating using magnetron sputtering or e-beam evaporation to form a uniform, optically opaque ground plane. The active top layer—graphene—is then incorporated via chemical vapor deposition (CVD) and subsequently transferred from a copper growth foil onto the FR-4 surface. A PMMA-assisted wet-transfer process, followed by gentle solvent cleaning, is used to maintain film continuity while minimizing cracks, wrinkles, and polymer residues [24–26].

The metasurface pattern is defined using standard photolithography, in which a spin-coated positive photoresist is exposed to UV light and developed. The unprotected graphene regions are then etched using oxygen plasma, a well-established method that offers high-resolution, low-damage patterning appropriate for THz devices [27–29]. For electrical interfacing or biasing, Ti/Au contact pads may be deposited and patterned using a lift-off process. Overall, this fabrication workflow provides accurate dimensional control, high repeatability, and full compatibility with commercial FR-4 processing and PCB manufacturing standards [30–32].

3.2. Alternative Additive Manufacturing:

For scalable and cost-efficient fabrication, electroless gold deposition may be used as an alternative to vacuum-based metallization. In this method, a thin metallic seed layer is first applied, followed by chemical gold plating, enabling uniform and well-adhered metal coverage on polymer substrates such as FR-4 [33–35]. After the graphene transfer, the metasurface can be patterned using non-photolithographic techniques, such as inkjet-printed resist masking followed by plasma etching, or femtosecond-laser direct writing (FsLDW). These additive or mask-free methods can achieve submicrometer resolution and have been successfully demonstrated in the fabrication of THz metasurfaces and flexible optoelectronic components [36–38]. This streamlined process reduces chemical usage and is suitable for large-area or roll-to-roll manufacturing.

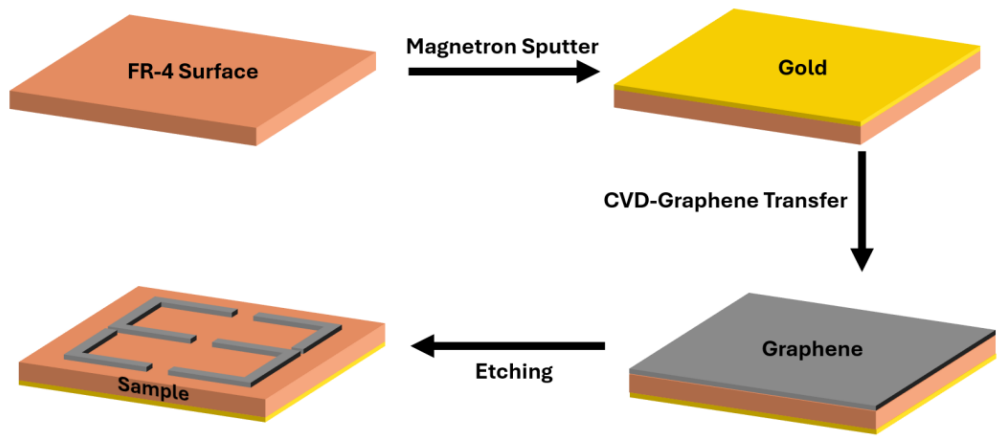

Fig 2: Overview of a possible process sequence for fabricating the proposed absorber

3.3. Final Processing and Assembly

All post-processing steps are maintained below 150 °C to avoid thermal damage to the FR-4 substrate. Final finishing procedures include residue removal, electrical isolation of the gold ground plane edges, and optional passivation of non-active regions. For RI sensing applications, an analyte containment frame or shallow microcavity may be added above the graphene layer. This streamlined device architecture—consisting of a single patterned graphene sheet, an FR-4 dielectric spacer, and a continuous gold ground plane—offers structural simplicity, low manufacturing cost, and strong compatibility with standard thin-film and PCB fabrication workflows [30–32].

## 4. Parametric analysis of the structure:

The parametric study provides insight into the geometric tuning mechanism. As shown in Fig. 3(a), the primary gap g is swept from 0.1 to 0.9 µm in increments of 0.2 µm. Increasing g reduces overall capacitance, resulting in a blue shift of the resonance frequency, with the highest absorption achieved at g = 0.5 µm. Conversely, reducing the auxiliary slit p increases capacitive loading, producing a slight red shift while maintaining stable absorption performance. These monotonic and predictable trends enable frequency tuning without modifying the unit cell footprint, as illustrated in Fig. 3(b). Material-level tunability through graphene parameters also proves effective. Figure 4(a) shows that an optimum absorption response occurs at a relaxation time of t = 0.2 ns, and increasing the relaxation time leads to a narrower absorption bandwidth, indicating potential enhancement of the Q-factor. Additionally, a parametric sweep of the chemical potential (Fermi level, $E_F$) from 0.3 to 0.7 eV is performed. A higher chemical potential shifts the resonance toward lower frequencies, as depicted in Fig. 4(b), demonstrating electronic post-fabrication control.

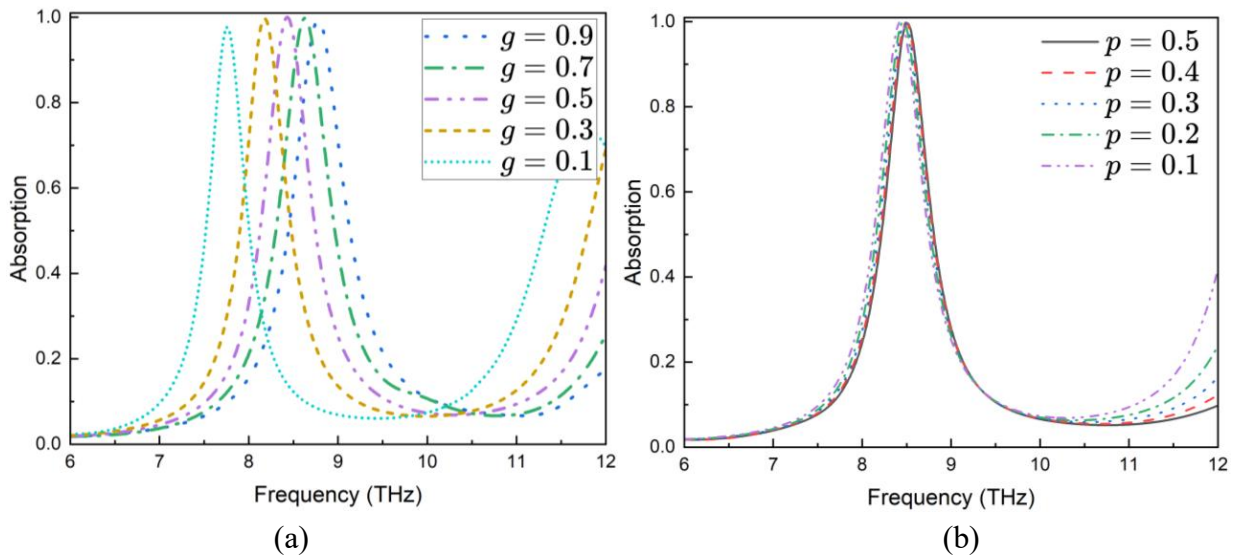

(a) (b)

Fig 3: The shifts of absorption peak with different geometric parameters (a) width of big gap, $g$ (b) width of small gap, $p$

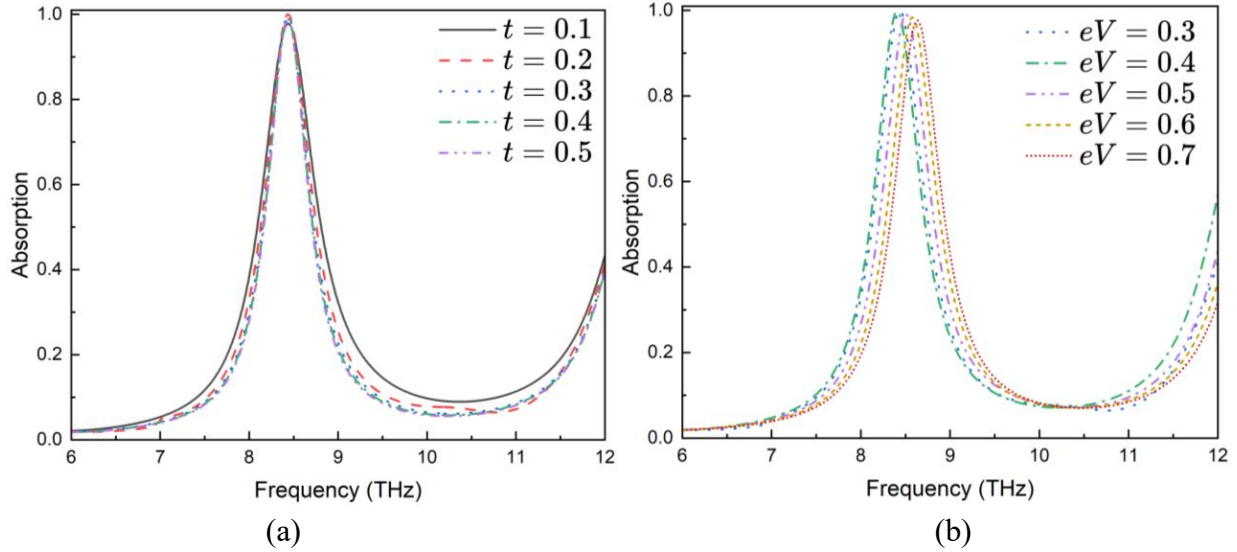

Fig 4: The shifts of absorption peak with different material properties of graphene (a) relaxation time, *t* (b) chemical potential, *eV*

## 5. Analysis of the proposed metamaterial absorber:

Absorption was obtained from the simulated S-parameters using $A(\omega) = 1 - |S_{11}|^2 - |S_{21}|^2$. Under normal incidence and periodic (Floquet) boundary conditions, the unit cell exhibits a single dominant absorption peak, as shown in Fig. 5(a–b). The resonance occurs at approximately 8.436 THz with a 208 GHz bandwidth. This behavior aligns with near-perfect absorption enabled by impedance matching to free space, confirmed by the retrieved normalized impedance where Re{z} ≈ 1 and Im{z} ≈ 0 at the peak frequency. Figure 6 further illustrates this result, showing that the real part of the impedance converges toward unity while the imaginary part approaches zero at resonance, confirming near-ideal impedance matching and consequently near-unity absorption.

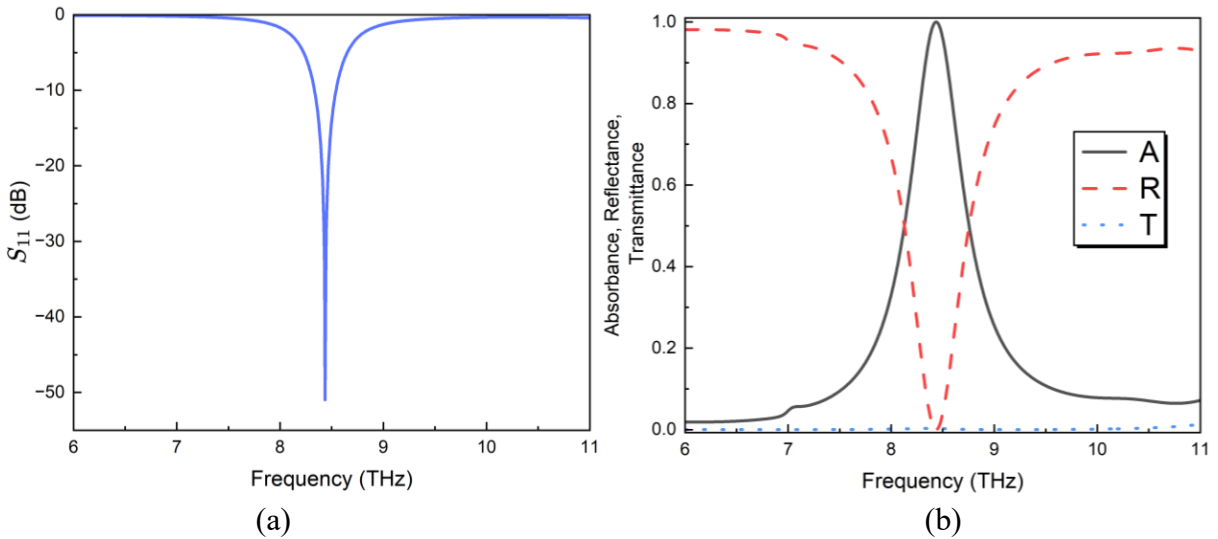

Fig 5: (a) Reflection co-efficient (b) Absorbance, reflectance and transmittance

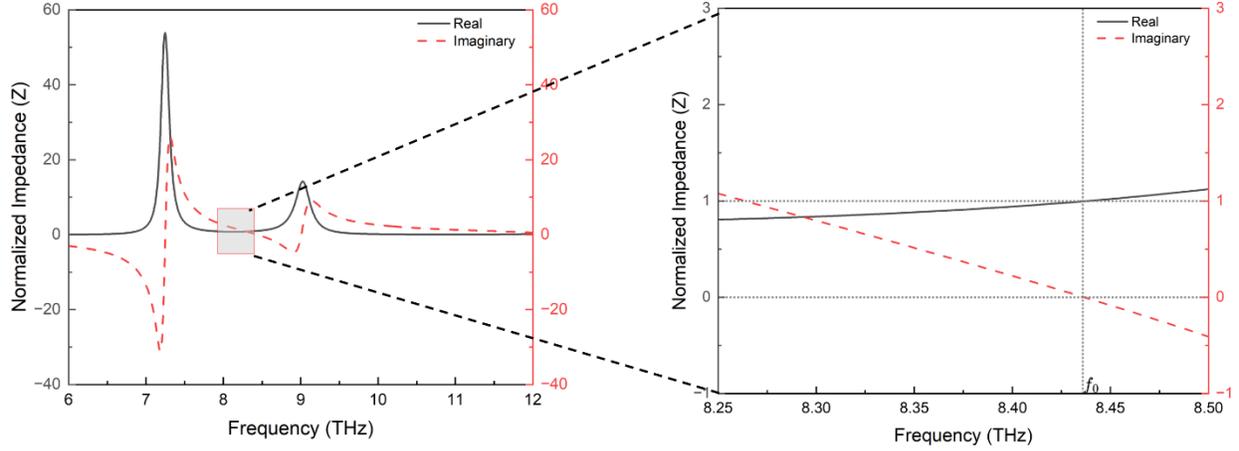

Fig 6: Real and imaginary components of the normalized impedance of the MTM sensor unit cell

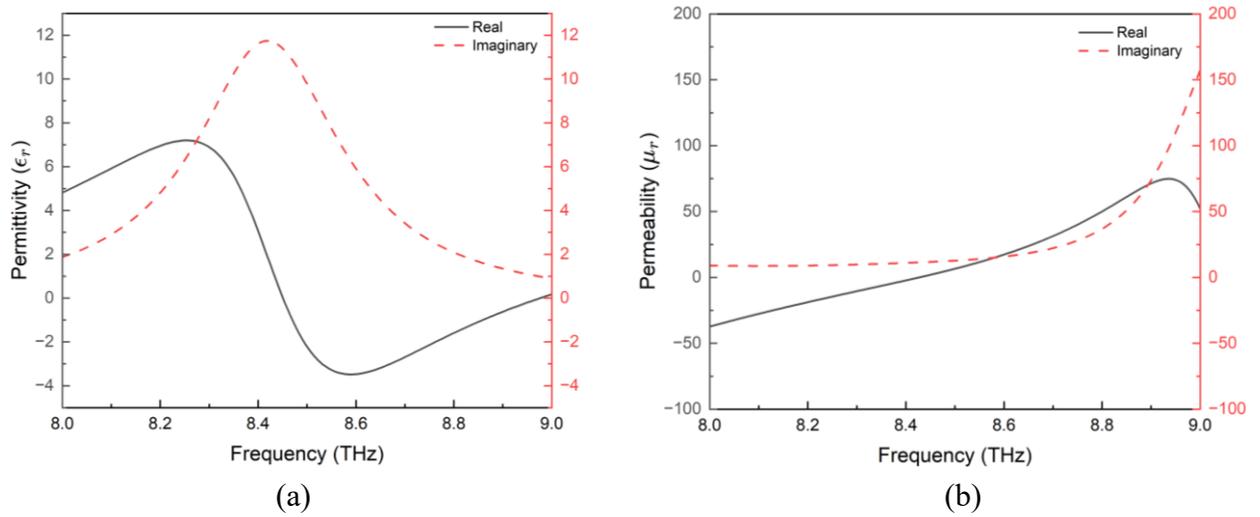

(a)            (b)

Fig 7: The relative permittivity and permeability of the unit cell (a) complex permittivity, $\varepsilon_r$ (b) complex permeability, $\mu_r$

The retrieval of effective electromagnetic parameters provides further insight into the electric–magnetic balance responsible for impedance matching. Both the real and imaginary components of the effective permittivity and permeability exhibit coupled dispersive behavior near the resonance, consistent with a strong electric response arising from the capacitive gaps and a magnetic response generated by the current loop between the patterned graphene layer and the continuous Au ground plane, as illustrated in Fig. 7.

For a metamaterial slab of thickness d and free-space wavenumber k₀ = 2πf / c, where f is the operating frequency and c is the speed of light, the effective permittivity and permeability can be derived from $S_{11}$, as expressed in [22],

$$\varepsilon_{\text{eff}} = 1 + \frac{2jS_{11} - 1}{k_0 d S_{11} + 1} \tag{1}$$

$$\mu_{\text{eff}} = 1 + \frac{2jS_{11} + 1}{k_0 d S_{11} - 1} \tag{2}$$

$$Z = \frac{Z_{\text{eff}}}{Z_0} = \sqrt{\frac{\mu_r}{\varepsilon_r}} \qquad (3)$$

Here, Z denotes the normalized impedance. Perfect absorption is achieved when Z = 1, meaning that the metamaterial's effective permittivity and permeability are balanced so that its input impedance matches that of free space. This matching condition generally requires simultaneous electric and magnetic resonances; if either response is absent or insufficient, significant impedance mismatch persists and perfect absorption cannot be obtained.

The retrieved effective parameters validate this mechanism, as shown in Fig. 7. At the resonant frequency, the relative permeability is approximately 0.867 + 11.637j, while the relative permittivity is −0.289 + 11.29j. The similarity in magnitude between these complex quantities leads to an impedance value close to unity, resulting in proper free-space matching. The large imaginary parts reflect strong electric and magnetic loss contributions, allowing efficient dissipation of incident energy and high absorption. Furthermore, the negative real part of ε_r at resonance indicates a left-handed (negative-index) response, which is associated with phenomena such as negative refraction and can further enhance absorber behavior [15,19].

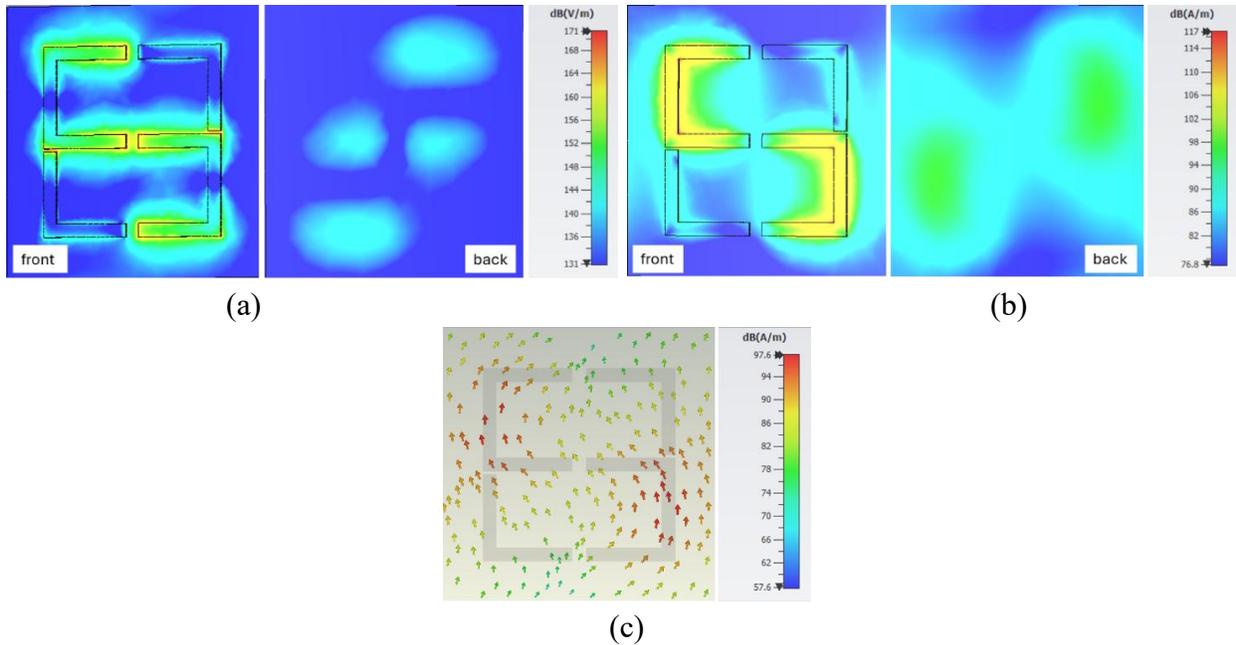

Fig 8: (a) Electric field distribution of the structure. (b) Magnetic field distribution of the structure. (c) The surface current.

A detailed field analysis was performed at the resonant frequency of 8.436 THz. Figure 8(a–b) shows the simulated electric and magnetic field distributions, providing physical insight into the operating mechanism of the proposed MM absorber. The resonance is primarily driven by the region formed between the C-shaped resonator segments and the auxiliary L-shaped split, as indicated by the strong electric-field concentration in Fig. 8(a). The field is tightly confined within these gaps, demonstrating that the patterned metasurface and dielectric spacer effectively trap incident electromagnetic energy, enabling near-unity absorption. To further clarify the underlying resonance mechanism, the magnetic-field magnitude |H| at 8.436 THz is shown in Fig. 8(b), and the corresponding surface-current distribution at 8.346 THz is plotted in Fig. 8(c). The induced ground-plane current forms a closed loop, confirming the excitation of a magnetic-dipole resonance. This coupling between strong capacitive electric-field confinement and magnetic-dipole current circulation explains the efficient energy dissipation and the high absorption peak.

## 6. Dependency on polarization and incident angles:

As noted earlier, the polarization and incident-angle responses of the proposed MM absorber were evaluated. The angular–polarization maps in Fig. 9 demonstrate that the unit cell supports a strong, magnetically driven TM-mode resonance near 8.436 THz, which remains narrowband and stable from $\theta = 0°$ to $75°$ with minimal frequency shift. High absorption levels are also maintained under polarization rotation from $\phi = 0°$ to $50°$, confirming both wide-angle and polarization tolerance. In comparison, Fig. 10 shows that the TE mode does not sustain a similar resonance due to the inherent structural asymmetry; however, near-unity absorption is still observed at $\theta = 0°$ when $\phi > 45°$. Overall, these results indicate that TM excitation offers the most reliable and trackable operating condition for sensing applications under realistic variations in incidence angle and polarization.

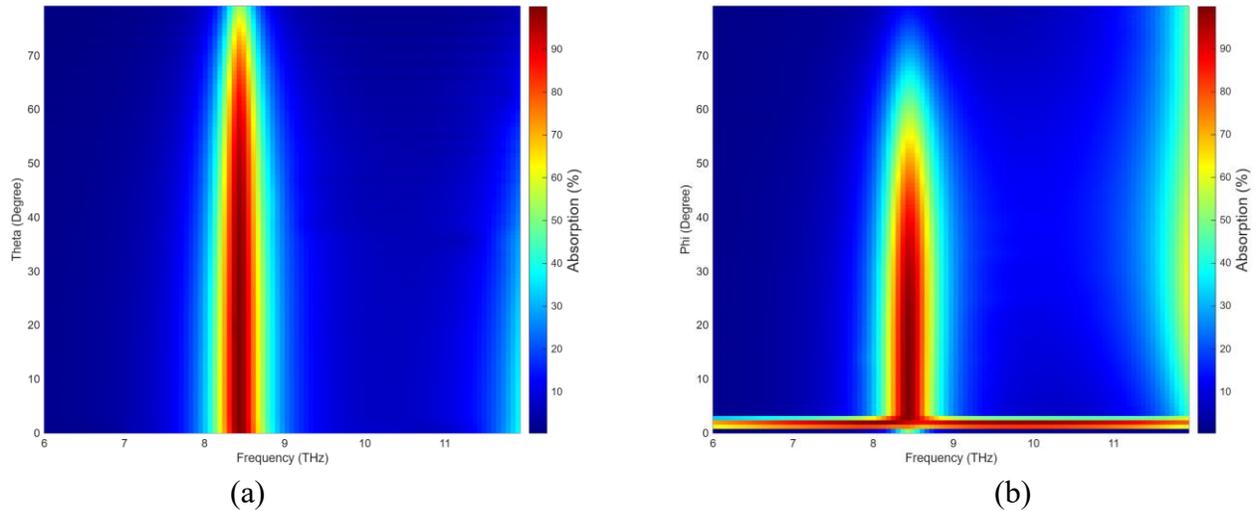

(a)                          (b)

Fig 9: Absorption spectra for TM mode at different (a) incident angle, $\theta$ (b) polarization angle, $\phi$

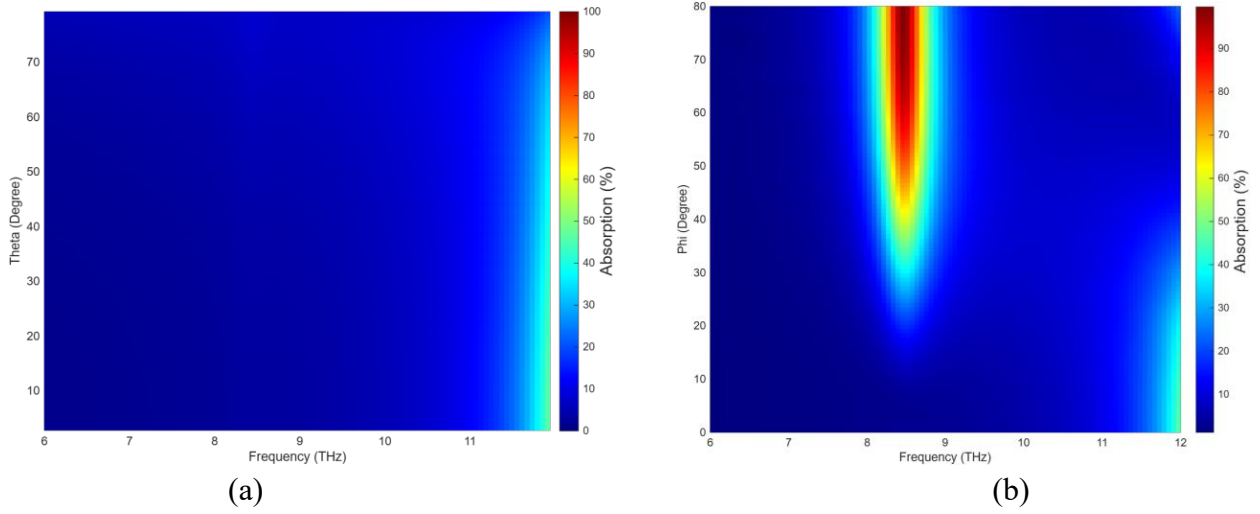

(a)                                                                  (b)

Fig. 10: Absorption spectra for TE mode at different (a) incident angle, $\theta$ (b) polarization angle, $\phi$

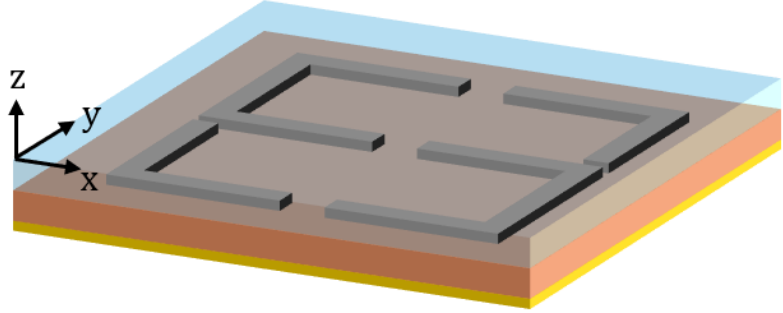

Fig 11: Proposed metamaterial absorber-based sensor with analyte layer

## 7. Design and analysis of proposed sensor performance:

For refractive-index (RI) sensing, an analyte superstrate layer is placed above the MM absorber, as illustrated in Fig. 11. The analyte RI is swept over a wide range from n = 1.0 to 2.0. The sensing performance is quantified using the following standard expression for sensitivity [22],

$$S = \frac{\Delta f}{\Delta n} \quad (4)$$

The figure of merit (FOM) is another key performance indicator for evaluating sensor quality, and it is defined as follows [22],

$$FOM = \frac{S}{FWHM} \quad (5)$$

In a metamaterial absorber, the quality factor (Q) characterizes the sharpness of the absorption resonance and is defined from the absorption spectrum as $Q = f_0 / FWHM$, where a higher Q corresponds to a narrower, more selective, and lower-loss peak. As the surrounding refractive index n increases, the resonance frequency shifts monotonically toward lower values. A linear fit of $f_0$ versus n yields a sensitivity of S = 1.698 THz/RIU. With a measured linewidth of FWHM =

0.402 THz, the resulting figure of merit (FOM) is 4.22 RIU$^{-1}$, and the associated quality factor is Q = 20.98. As shown in Fig. 13, the linear regression confirms a direct proportional relationship between the resonance frequency and the refractive index of the surrounding medium, expressed by the equation,

$$f_0 = -1.698n + 10.13 \tag{6}$$

Although the resonance exhibits a comparatively lower Q-factor and FOM, the sensor demonstrates significantly higher RI sensitivity (S), which is the most critical parameter for frequency-shift–based sensing. The reduced Q and FOM primarily result from the FR-4 substrate, whose relatively high dielectric and conductive losses at THz frequencies broaden the resonance response.

Figure 12(a–b) summarizes the absorption spectra, peak evolution across the RI sweep, and the corresponding linear calibration fit. The sensing range spans n = 1.0–1.2 for gases and porous media, n = 1.3–1.7 for oils and polymeric materials, and extends up to n = 2.0 for high-index compounds. This range also encompasses the biomedical sensing window (n = 1.30–1.39), relevant to many aqueous bio-analytes [19,22].

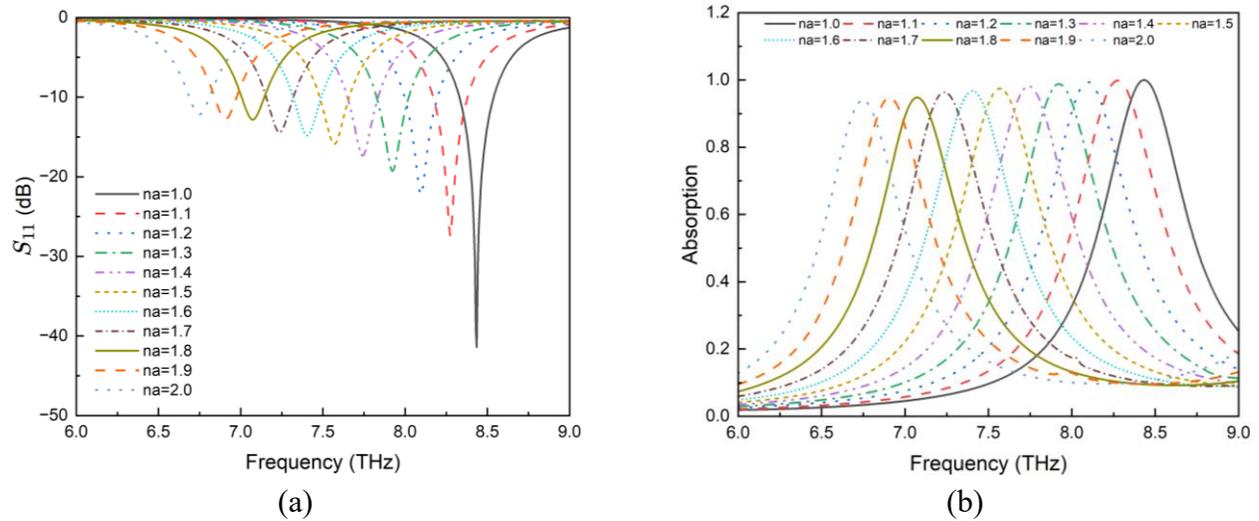

Fig 12: (a) S-parameters (b) absorption spectrum−variations due to refractive index changes in the analyte

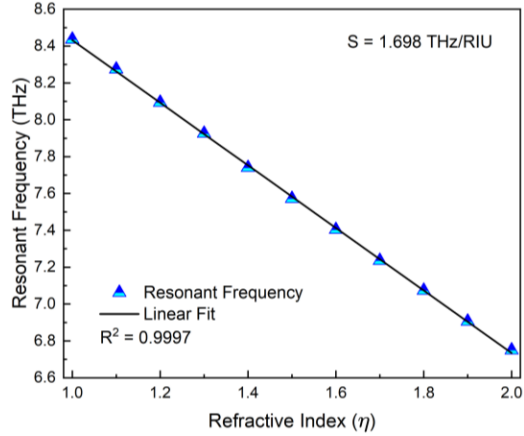

Fig 13: Plot of resonant frequencies versus refractive index

## 8. Equivalent circuit model of the sensor:

The graphene–FR-4–Au unit cell is modeled as a single resonant branch located on the top surface above a conductive ground plane. The continuous Au backplane behaves as an RF short at THz frequencies and therefore does not contribute an independent resonance; instead, the absorption response is governed entirely by the lumped reactive behavior of the patterned graphene layer, consistent with standard modeling practice for metal-backed metamaterial absorbers and sensors [42]. In the equivalent circuit, the graphene conduction path is represented by a series inductance $L_1$ and a resistive term $R_1$, which accounts for Ohmic and scattering losses associated with the graphene sheet impedance. The central split gaps give rise to the dominant capacitance $C_1$, while near-edge coupling between adjacent arms contributes an auxiliary capacitance $C_2$. As illustrated in Fig. 14(a), the resulting topology corresponds to a lossy series L–C resonator whose input impedance, referenced through the FR-4 substrate, reaches a reflection minimum when it approaches the free-space impedance [43,44].

The lumped-element values were initially estimated using closed-form expressions that relate capacitance and inductance to the geometry of the S-shaped resonator. For one vertical strip and its horizontal segment, the inductance is approximated using the thin-strip loop formulation,

$$L_1 = \mu_0 l_{eq} \left[ \ln\left(\frac{2 l_{eq}}{m}\right) - \frac{1}{2} \right]; \quad l_{eq} \approx 2(a - m) + (b - g) \tag{7}$$

where $\mu_0$ is the permeability of free space, m denotes the graphene strip width, a represents the vertical arm length, b is the horizontal overlap across the split, and g is the split-gap separation. The term l_eq accounts for the effective current path length along the top branch of the resonator [45,46].

The dominant split-gap capacitance associated with the central bar is estimated as

$$C_1 = \frac{\varepsilon_0 \varepsilon_{eff} m b}{g} \tag{8}$$

with $\varepsilon_0$ the vacuum permittivity and $\varepsilon_{eff} \approx (\varepsilon_r + \frac{n^2}{2})$ the effective permittivity seen by the fringing field [45,47].

The electric-field build-up between the two parallel vertical arms provides an additional capacitive contribution. The inter-arm coupling capacitance was estimated by a coplanar-strip expression,

$$C_2 = \frac{2\pi\varepsilon_0\varepsilon_{eff}l_c}{\ln\left(\frac{p+m}{p}\right)} \quad ; \quad l_c \approx a - 2m, \tag{9}$$

where $p$ is the edge-to-edge spacing between the two arms and $l_c$ is their parallel overlap length [5–7].

In the circuit model, the split-gap capacitance $C_1$ together with the inter-arm coupling capacitance $C_2$ form the effective capacitance

$$C_{eq} = C_1 + C_2 \tag{10}$$

and the absorption resonance of the grounded metasurface is governed by the one-port series-LC condition

$$f_0 = \frac{1}{2\pi\sqrt{L_1 C_{eq}}} \tag{11}$$

which sets the reflection minimum for a metal-backed absorber [43,45].

The element values were then fine-tuned until the equivalent circuit's reflection coefficient closely matched the full-wave results from CST Studio Suite−3D Simulation.

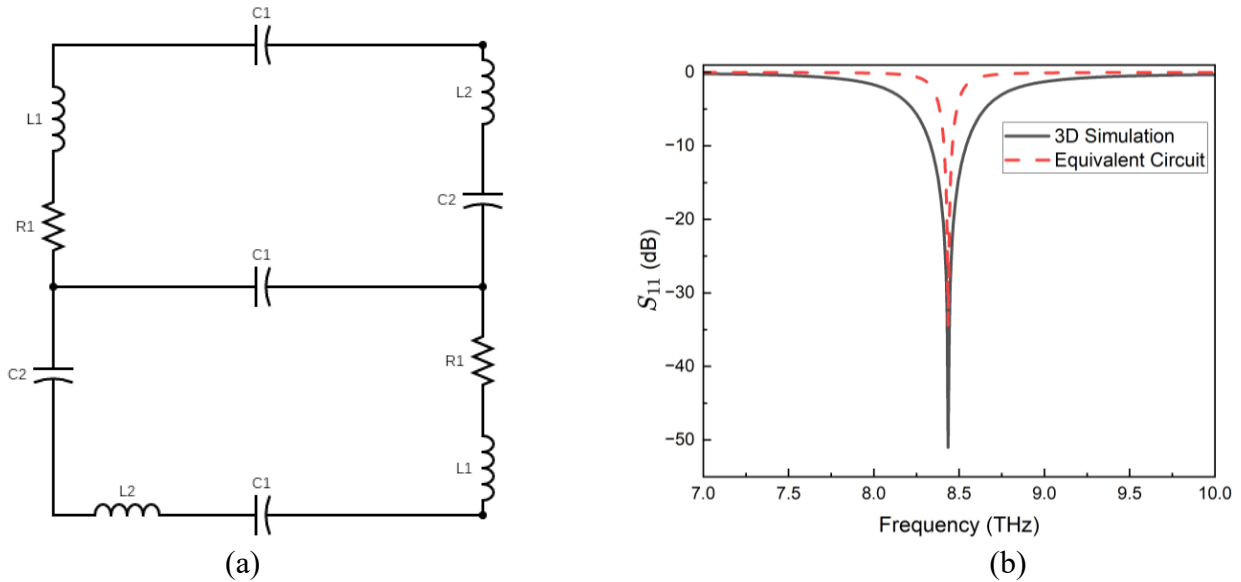

(a) (b)
Fig 14: (a) Equivalent Circuit Design (b) Reflection coefficient spectra simulated by CST software.

## 9. Comparative study with previously reported literature:

Table 1 summarizes a comparison of recent THz metamaterial (MM)–based sensors. Floating metasurface designs report a high responsivity of 532 GHz/RIU across n = 1.00–1.47, although Q and FOM metrics are not provided [48]. InSb-based perfect absorbers support dual temperature–RI sensing and demonstrate a relatively large shift of 1043 GHz/RIU with a moderate Q = 53.24 in the n = 1.00–1.10 range [49]. A multiband SRR biosensor achieves a sensitivity of 642.5 GHz/RIU within n = 1.33–1.40 [50]. Meanwhile, a lithium-niobate BIC metasurface attains an ultra-high Q = $6.56\times10^4$ over n = 1.20–1.40, illustrating the upper limit of resonator sharpness, although FOM data is not reported [51]. Dual-band absorbers designed for skin-cancer diagnostics operate over n = 1.30–1.40, but show lower sensitivities in the 100–398 GHz/RIU range [52,53]. Microfluidic absorber platforms, which include integrated analyte handling, yield 785 GHz/RIU with Q = 23.5 over n = 1.00–1.40 [54]. Recently, triple-band and polarization-insensitive absorbers have demonstrated improved linewidth characteristics with FOM values between 6.95–7.60 and sensitivities of 374–1500 GHz/RIU, though within narrower aqueous-index windows (typically n = 1.30–1.34 or 1.34–1.40) [55,56]. A pesticide-detection absorber covering n = 1.00–1.25 offers a general-purpose reference but with relatively low performance (208 GHz/RIU, FOM = 2.53) [45]. In comparison, the present work delivers the highest sensitivity among surveyed absorber-based THz sensors (1698 GHz/RIU), along with a competitive Q-factor (20.98) and an improved FOM (4.22) relative to several reported platforms. Additionally, it provides the widest validated RI sensing range (n = 1.0–2.0) without requiring extreme fabrication symmetry as in BIC-based systems [51] or thermally triggered phase transitions used in $VO_2$-based architectures [53].

**Table 1:** Sensing performance comparison of proposed THz MM sensor with previously reported sensors

| References | Max. Sensitivity (GHz/RIU) | Max Q-factor | FOM | RI range | Year |
|---|---|---|---|---|---|
| [48] | 532 | - | - | 1.00−1.47 | 2021 |
| [49] | 1043 | 53.24 | - | 1.00−1.10 | 2021 |
| [50] | 642.5 | - | - | 1.33−1.40 | 2023 |
| [51] | 947 | $6.56\times10^4$ | - | 1.20−1.40 | 2023 |
| [52] | 100 | - | - | 1.30−1.40 | 2024 |
| [53] | 398 | - | - | 1.30−1.40 | 2024 |
| [54] | 785 | 23.50 | - | 1.00−1.40 | 2025 |
| [45] | 208 | 22.46 | 2.53 | 1.00−1.25 | 2025 |
| [55] | 1500 | 39.13 | 6.95 | 1.34–1.40 | 2025 |
| [56] | 374 | 28.26 | 7.60 | 1.30–1.34 | 2025 |
| **This work** | **1698** | **20.98** | **4.22** | **1.0−2.0** | **-** |

## 10. Conclusions:

The proposed graphene-enabled THz metamaterial absorber offers a compact and electrically tunable platform for refractive-index sensing across an exceptionally wide operational window of

n = 1.0–2.0. Full-wave electromagnetic analysis confirms near-perfect, impedance-matched absorption at 8.436 THz with FWHM = 0.402 THz and Q = 20.98, corroborated by normalized impedance retrieval and field-distribution analysis showing a coupled capacitive–inductive resonance localized at the graphene gaps and closed through the gold ground plane. The resonance frequency exhibits a linear, monotonic red-shift with increasing analyte RI, resulting in a high sensitivity of 1698 GHz/RIU and FOM = 4.22 $RIU^{-1}$, enabling a single device to cover gas analytes, oils/polymers, and the biomedical aqueous index range. Electrical tunability through graphene's chemical potential, together with strong near-field confinement, allows practical post-fabrication trimming of resonance frequency and absorption magnitude while retaining stability under variations in incidence angle and polarization. Future work will focus on experimental fabrication and characterization, determining limits of detection under realistic noise and drift conditions, and exploring microfluidic integration and surface functionalization for selective biosensing.